%% file: main.tex
\documentclass{article}

\usepackage{arxiv}

\usepackage[utf8]{inputenc} 
\usepackage[T1]{fontenc}    
\usepackage{hyperref}       
\usepackage{url}            
\usepackage{booktabs}       
\usepackage{amsfonts}       
\usepackage{nicefrac}       
\usepackage{microtype}      
\usepackage{lipsum}
\usepackage{graphicx}
\graphicspath{ {./images/} }
\usepackage{graphicx}
\usepackage{multicol,multirow}
\usepackage{amsmath,amssymb,amsfonts}
\usepackage{mathrsfs}
\usepackage{amsthm}
\usepackage{rotating}
\usepackage{appendix}
\usepackage{ifpdf}
\usepackage[T1]{fontenc}
\usepackage{newtxtext}
\usepackage{newtxmath}
\usepackage{textcomp}
\usepackage{xcolor}
\usepackage{natbib}
\definecolor{orange}{HTML}{cc7a00}	

\usepackage{lipsum}
\usepackage{nicefrac}
\usepackage{float}

\theoremstyle{definition}

\numberwithin{equation}{section}

\input{commands}

\title{Mean flow data assimilation using physics-constrained Graph Neural Networks}

\author{
 Michele Quattromini \\
  Dipartimento di Meccanica, Matematica e Management\\
  Politecnico di Bari\\
  Bari, 70126 Italy  \\
  LISN-CNRS\\
  Université Paris-Saclay\\
  Orsay, 91440 France\\
  \texttt{michele.quattromini@poliba.it} \\
   \And
 Michele Alessandro Bucci \\
  Digital Sciences \& Technologies Department\\
  SafranTech\\
  Magny-Les-Hameaux, 78114 France \\
  \And
 Stefania Cherubini \\
  Dipartimento di Meccanica, Matematica e Management\\
  Politecnico di Bari\\
  Bari, 70126 Italy \\
  \AND
  Onofrio Semeraro \\
  LISN-CNRS\\
  Université Paris-Saclay\\
  Orsay, 91440 France\\
}

\begin{document}
\maketitle
\begin{abstract}Despite their widespread use, purely data-driven methods often suffer from overfitting, lack of physical consistency, and high data dependency, particularly when physical constraints are not incorporated. This study introduces a novel data assimilation approach that integrates Graph Neural Networks (GNNs) with optimisation techniques to enhance the accuracy of mean flow reconstruction, using Reynolds-Averaged Navier--Stokes (RANS) equations as a baseline. The method leverages the adjoint approach, incorporating RANS-derived gradients as optimisation terms during GNN training, ensuring that the learned model adheres to physical laws and maintains consistency. Additionally, the GNN framework is well-suited for handling unstructured data, which is common in the complex geometries encountered in Computational Fluid Dynamics (CFD). The GNN is interfaced with the Finite Element Method (FEM) for numerical simulations, enabling accurate modelling in unstructured domains. We consider the reconstruction of mean flow past bluff bodies at low Reynolds numbers as a test case, addressing tasks such as sparse data recovery, denoising, and inpainting of missing flow data. The key strengths of the approach lie in its integration of physical constraints into the GNN training process, leading to accurate predictions with limited data, making it particularly valuable when data are scarce or corrupted. Results demonstrate significant improvements in the accuracy of mean flow reconstructions, even with limited training data, compared to analogous purely data-driven models.
\end{abstract}



\input{TexFiles/3_intro}
\input{TexFiles/4_data_assim}
\input{TexFiles/5_numerics}
\input{TexFiles/6_ML}
\input{TexFiles/7_methodology}
\input{TexFiles/8_results}
\input{TexFiles/9_conclusion}




\begin{appendix}
\input{TexFiles/10_appA}
\end{appendix}

\bibliographystyle{abbrv} 
\bibliography{references}

\end{document}

%% file: commands.tex
\newcommand{\vdir}{\mathbf{u}}
\newcommand{\vpred}{\hat{\mean\vdir}}
\newcommand{\ppred}{\hat{\mean p}}

\newcommand{\pdir}{p}
\newcommand{\ff}{\mathbf{f}}
\newcommand{\fpred}{\hat{\ff}}

\newcommand{\mean}[1]{\overline{#1}}
\newcommand{\fluct}[1]{#1'}
\newcommand{\invRe}{\dfrac{1}{Re}}

%% file: TexFiles/3_intro.tex
\section{Introduction}\label{sec:intro}

In recent years, the integration of Machine Learning (ML) algorithms into Computational Fluid Dynamics (CFD) has experienced significant growth, driven by the increasing efficiency of ML models in processing large datasets and their impressive capabilities in inference and prediction. The literature is already rich with various effective approaches for combining ML algorithms with CFD, as highlighted in the review works by Brunton \emph{et al.} \cite{brunton2020machine} and Vinuesa \emph{et al.} \cite{vinuesa2022enhancing}. These applications have found a natural domain in turbulence modelling problems, ranging from the closure of Reynolds-Averaged Navier-Stokes (RANS) equations to wall-modelled Large Eddy Simulations \citep{lapeyre2019training, zhou2021wall, bae2022scientific}. Among the many research efforts addressing these challenges, Ling \& Templeton \cite{LingTempleton2015} applied classification methods to identify regions of uncertainty where the closure term in RANS might fail, while other approaches leverage baseline models such as the Spalart-Allmaras closure \citep{singh2016using}, physics-informed neural networks (PINNs) \citep{eivazi2022physics}, regression methods \citep{schmelzer2019machine}, decision trees \citep{duraisamy2019turbulence}, ensemble methods \citep{mcconkey2022generalizability}, Tensor-Basis Neural Networks \citep{ling2016reynolds, cai2024revisiting}, and genetic programming \citep{weatheritt2016novel, zhao2020rans}. For a broader overview, we refer to references \cite{duraisamy2019turbulence}, \cite{beck2021perspective}, and \citep{cinnella2024data}, where thorough reviews of ML techniques applied to the subject are provided.

On the other hand, alternative approaches are based on data assimilation. In this framework, diverse observations are integrated with the governing dynamical laws of a system to optimally estimate physical variables. Starting from a background state and incorporating imperfect observational data, it yields the best possible estimate of the system’s true state by accounting for the statistical reliability of each data source. These methodologies are routinely employed in meteorology, where accurate weather forecasting is achieved by combining satellite imagery, meteorological measurements, and numerical models \cite{titaud2010assimilation}, \cite{navon2009data}. Two main approaches to data assimilation can generally be identified: stochastic estimation, which relies on probabilistic methods such as Kalman filtering, and variational data assimilation. In the context of fluid mechanics, recent studies employing stochastic estimation include \cite{meldi2017reduced, cheng2019background}. Variational assimilation, and in particular the four-dimensional variational assimilation (4D-Var) methods have been applied in \cite{cordier2013identification} for turbulence model identification, and in a similar spirit in \cite{foures2014data}. The success of these applications in fluid mechanics is not surprising as variational methods based on adjoint formulations are commonly applied across a broad range of problems. These include sensitivity analysis in stability studies \citep{giannetti2007structural, marquet2008sensitivity, luchini2009structural}, optimisation \cite{schmid1994optimal, cherubini2010optimal, magri2019adjoint, giannotta2022minimal}, receptivity analysis \cite{giannetti2006leading}, and flow control \citep{cherubini2013nonlinear, semeraro2013riccati}. A thorough overview of these applications is provided by Luchini and Bottaro \cite{luchini2014adjoint}. Returning to the topic of mean flow reconstruction, Foures et al. \cite{foures2014data} proposed a variational data assimilation method for reconstructing the mean flow field from partial measurements using the forced RANS equations. Their approach minimizes the discrepancy between experimental data and numerical solutions by identifying an optimal forcing term that effectively represents the unknown Reynolds stresses. This is achieved through a direct-adjoint looping procedure and the closure model is interpreted as an optimal control variable informed by the adjoint field.

Building on this, recent efforts combining data assimilation with neural network (NN) modelling have been proposed in references \citep{volpiani2021machine, strofer2021end, patel2024turbulence}. Following this philosophy, in this work, we propose a novel approach that integrates Graph Neural Networks (GNNs) with Reynolds-Averaged Navier-Stokes (RANS) equations to enhance the training process. The resulting method is referred to as the Physics-Constrained Graph Neural Network (PhyCo-GNN). Specifically, the forcing term in the RANS equations is modelled using a GNN, where the network is informed by gradients obtained via auto-differentiation and analytical gradients derived from the adjoint equations during the iterative process. By embedding the RANS closure term into the optimisation framework via the adjoint method, we leverage adjoint methods to efficiently compute gradients necessary for optimisation. This ensures that the gradients used during the GNN training are informed by a deterministic physical model, rather than relying solely on available data, thus ensuring physical consistency. A key feature of this approach is the use of GNNs \citep{hamilton2020graph}, which are characterised by complex, multi-connected nodes that can be naturally adapted to unstructured meshes. In a GNN, convolution is performed by aggregating information from neighbouring nodes, overcoming the limitations of geometry typically encountered in Convolutional Neural Networks (CNNs). Additionally, GNNs exhibit superior generalisation capabilities compared to standard models \citep{sanchez2018graph}, are differentiable, and enable direct learning of operators through discrete stencils \citep{shukla2022scalable}. Due to these advantages, GNNs have recently gained attention in fluid mechanics. A comprehensive review by Lino \emph{et al.} \cite{lino2023current} and examples from \cite{toshev2023learning} and \cite{dupuy2023modeling}, where GNNs are used to model wall shear stress in LES simulations, highlight their potential.
In Quattromini \emph{et al.} \cite{quattromini2023operator}, a GNN is used to predict the closure term in RANS equations at low Reynolds numbers, using a supervised learning strategy. In that setting, the NN model receives the mean flow field as input and is trained to regress the corresponding forcing term, with both quantities derived from high-fidelity DNS data. The approach is designed to map the mean flow to the forcing term in a purely data-driven setting.
In contrast, the present work incorporates this GNN architecture as a pre-trained module within a broader physics-constrained optimisation framework. The GNN model is here repurposed to act as a prior for a data assimilation scheme governed by the RANS equations. The key novelty lies in the integration of the GNN into an adjoint-based inverse problem, where the NN model is trained not to only replicate the DNS-derived forcing term, but also to minimise discrepancies in the mean flow field reconstruction through gradient-based optimisation, informed by the governing physics.
The remainder of this article is structured as follows. The mathematical framework is described in Sec.~\ref{sec:DA}, where the physical baseline model for the numerical simulations is detailed in Sec.~\ref{sec:governing_equations}, and the adjoint optimisation method is presented in Sec.~\ref{subsec:adjoint_methods}. Sec.~\ref{sec:gnn_overview} outlines the ML framework, detailing the GNN architecture (Sec.~\ref{subsec:mp_algorithm}), the dataset preprocessing (Sec.~\ref{subsec:data_structure}), and the training algorithm (Sec.~\ref{subsec:training_process_algorithm}). We then proceed to present the PhyCo-GNN approach in Sec.~\ref{sec:methodology}. Results are presented in Sec.~\ref{sec:results}, where mean flow reconstruction is shown for several test cases, including sparse data recovery, denoising, and inpainting of missing flow data. Conclusions are drawn in Sec.~\ref{sec:conclusion}.

%% file: TexFiles/4_data_assim.tex
\section{Mean Flow Reconstruction Using Data Assimilation}\label{sec:DA}

\subsection{Governing Equations}\label{sec:governing_equations}
This study focuses on two-dimensional (2D) incompressible fluid flows around bluff bodies in unsteady regimes. We begin with the Navier-Stokes (NS) equations for incompressible flows. Let $\mathbf{x}=(x,y)$ denote the spatial Cartesian coordinates. The velocity field ${\vdir}(\mathbf{x}, t)$ and pressure field $p(\mathbf{x}, t)$ follow these dynamics: 
\begin{subequations}
\begin{align}
\dfrac{\partial \vdir}{\partial t} +\left(\vdir\cdot\nabla\right) \vdir &= 
-\nabla\pdir +\invRe\nabla^2\vdir \\
\nabla\cdot\vdir &= 0. 
\end{align}
\label{eq:unsteadyNS}
\end{subequations}
The above equations are made dimensionless by introducing the characteristic length scale $D$ (\emph{e.g.}, the diameter of the circumscribed bluff body circumference), the velocity $U_{\infty}$ of the incoming flow, and $\rho U_{\infty}^2$ as the reference pressure. The Reynolds number is defined as $Re = \nicefrac{U_{\infty} D}{\nu}$, where $\nu$ represents the kinematic viscosity. This number represents the ratio between the inertial forces and the viscous forces in the fluid flow and is used to characterise the flow regime, indicating whether the flow is laminar or turbulent, depending on the specific case. In this study, we focus on unsteady nonlinear flows that settle on limit cycles or quasi-periodic regimes, a configuration typically occurring when supercritical flows have not yet become turbulent.

The NS equations can be computationally intensive. Hence, various approximate models are used based on the required accuracy. In this study, we adopt the Reynolds-averaged Navier-Stokes (RANS) model, a time-averaged formulation of the NS equations. By introducing the Reynolds decomposition: 
\begin{equation}
\vdir(\mathbf{x},t) = \mean\vdir(\mathbf{x}) +\fluct\vdir(\mathbf{x},t), \label{eq:reydeco} 
\end{equation}
the unsteady velocity field $\vdir=(u,v)^T$ is split into an ensemble-averaged velocity field $\mean{\vdir}=(\mean{u},\mean{v})^T$ and a fluctuating component $\fluct\vdir=({\fluct u},{\fluct v})^T$. Formally, this decomposition allows any unsteady flow to be expressed as a sum of a steady mean flow and an unsteady fluctuating part. Plugging the Reynolds decomposition (Eq.~\ref{eq:reydeco}) into the NS equations (Eq.~\ref{eq:unsteadyNS}) and ensemble-averaging results in: 
\begin{subequations}
\begin{align}
\left(\mean\vdir\cdot\nabla\right) \mean\vdir &= 
-\nabla\mean\pdir
+\invRe\nabla^2\mean\vdir 
+\ff\\
\nabla\cdot\mean\vdir &= 0,
\end{align}
\label{Eq:RANS}
\end{subequations}
where $\mean{p}$ is the mean pressure field. These equations are completed with a set of boundary conditions, detailed in Sec.~\ref{sec:numerical}, together with a description of the numerical setup.

In this context, the Reynolds stress tensor $\ff$ acts as a closure term for the underdetermined system of nonlinear equations in Eq.~\ref{Eq:RANS}. Ideally, $\ff$ can be directly computed, when data are available, as:
\begin{equation}
\ff= -\nabla \cdot (\overline{\fluct\vdir \fluct\vdir}).\label{eq:forcing}
\end{equation}
In practice, mathematically computing $\ff$ requires either Direct Numerical Simulation (DNS) or sufficiently long sampling of experimental or numerical data to accurately capture the statistical properties of the fluctuating components $\fluct\vdir$. Unlike the mean flow $\mean\vdir$, the fluctuating components are inherently statistical in nature, and their proper representation depends on achieving statistical convergence rather than full time-resolved data. This is known as the closure problem. Several approximations, like the Boussinesq hypothesis — \emph{e.g.} $k$-$\epsilon$, $k$-$\omega$, Spalart-Allmaras models \citep{wilcox1998turbulence} — or more complex models such as the explicit algebraic Reynolds stress model \citep{wallin2000explicit} and differential Reynolds stress models \citep{cecora2015differential}, can be introduced to address this problem. Nevertheless, in this work, we do not assume any modelling approximation of the forcing term $\ff$ but instead aim to compute it by combining neural-network modelling and data assimilation; a GNN model is trained to infer the Reynolds stress term $\ff$ from a given mean flow $\mean\vdir$ introduced as input. The Reynolds stress term $\ff$ is the output of the GNN and is assumed in this context as a control variable in the adjoint optimisation process (see Sec.\ref{sec:methodology}).

\begin{figure}[t]
\includegraphics[width=1\textwidth]{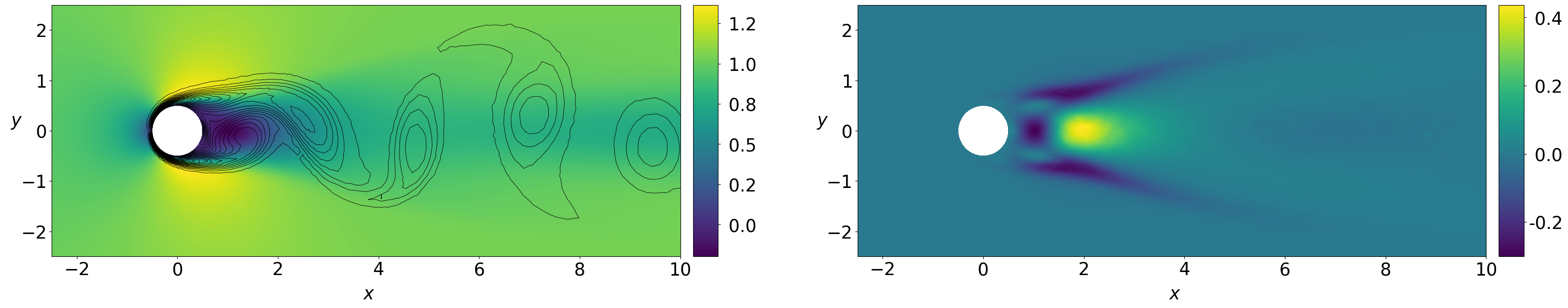}
\vspace{-0.5cm}
\caption{
$(a)$ Streamwise component of the mean flow, $\mean \vdir$, and vorticity isolines, $\omega = \nabla \times \vdir$, for the flow past a cylinder at $Re = 150$. $(b)$ For the same case, the streamwise component of the closure term, $\mathbf{f}$, is shown. In both cases, only a portion of the domain is displayed}\label{fig:mean_vort}
\begin{picture}(0,0)
\put(5, 130){$a)$}
\put(215, 130){$b)$} 
\end{picture}
\end{figure}

\subsection{Adjoint-based Data Assimilation}\label{subsec:adjoint_methods}
Among the various strategies available in the literature, data assimilation schemes can be formulated by casting optimisation loops based on the adjoint method. In this approach, a cost function is defined to be either maximised or minimised, alongside a control variable to be adjusted accordingly. To this end, we introduce a mapping that projects the mean flow field $\mean \vdir$ onto the measurement vector $\bar{\mathbf{m}}$ as follows
$$
\bar{\mathbf{m}}=\mathcal{M}{(\mean\vdir)}.
$$
The operator $\mathcal{M}$ defines the ground truth data used in the optimisation process. If $\mathcal{M} = I$, we consider the full-field $\mean \vdir$; otherwise, $\mathcal{M}$ may be designed to project the field onto a subset of measurements (e.g., sparse observations or an incomplete flow field, as in the inpainting scenario; see Sec.~\ref{sec:results}). Based on this definition, the cost function to be minimised is given by the discrepancy between the ground truth $\bar{\mathbf{m}}$ and the reconstructed measurements $\mathcal{M}(\vpred)$
\begin{equation}
    \mathcal{J}\left(\vpred\right) = \frac{1}{2} ||\bar{\mathbf{m}} - \mathcal{M}(\vpred)||_2^2,
    \label{eq:cost_function}
\end{equation} 
where $||\cdot||_2$ represents the $L^2$-norm, defined in association with the scalar product over the domain $\Omega$:
\begin{equation}
    \langle\mathbf{a}, \mathbf{b}\rangle = \int_\Omega \mathbf{a}\cdot\mathbf{b} \, d\Omega.
    \label{eq:scalar_product}
\end{equation}

We consider the RANS equations (Eq.~\ref{Eq:RANS}) as the baseline equation and choose the unknown forcing stress term $\ff$ as the control variable. This procedure is inspired by the study \cite{foures2014data}, where an optimisation loop is designed to reconstruct the mean flow $\mean \vdir$.\\ Since the control variable is the forcing stress term $\ff$, the cost function $\mathcal{J}$ (Eq.~\ref{eq:cost_function}) does not explicitly depend on it. Therefore, to relate the cost function $\mathcal{J}$ to the forcing stress tensor $\ff$, we need to define an augmented Lagrangian functional $\mathcal{L}$:
\begin{equation}
    \mathcal{L}\left(\ff, \vpred, \ppred, \vpred^\dag, \ppred^\dag\right) =
    \mathcal{J} \left(\vpred\right) - \langle\vpred^\dag, \vpred\cdot\nabla\vpred + \nabla\ppred - \invRe\nabla^2\vpred - \ff\rangle - \langle\ppred^\dag, \nabla\cdot\vpred\rangle
    \label{eq:lagrangian_augmented}
\end{equation} 
where $\langle \cdot, \cdot \rangle$ represents the spatial scalar product, as defined in Eq.~\ref{eq:scalar_product}. The augmented Lagrangian functional $\mathcal{L}$ allows us to represent the constrained optimisation problem as an unconstrained one by introducing two a-priori unknown variables, the Lagrangian multipliers $\vpred^\dag$ and $\ppred^\dag$. To minimise the functional defined in Eq.~\ref{eq:lagrangian_augmented}, its partial derivatives with respect to all independent variables of the problem must be set to zero.

Following this approach, nullifying the partial derivatives of Eq.~\ref{eq:lagrangian_augmented} with respect to the direct variables $\vpred$ and $\ppred$ yields the adjoint Navier-Stokes equations:
\begin{subequations}
    \begin{align}
        -\vpred\cdot\nabla\vpred^\dag + \vpred^\dag \cdot \nabla\vpred^T - \nabla\ppred^\dag - \invRe\nabla^2\vpred^\dag &= \dfrac{\partial\mathcal{J}}{\partial\vpred}\\
        \nabla\cdot\vpred^\dag &= 0 
    \end{align}
    \label{eq:adjoint_NS}
\end{subequations}
along with an appropriate set of boundary conditions, detailed in Sec.~\ref{sec:numerical}. It is worth noting that the adjoint NS equations (Eq.~\ref{eq:adjoint_NS}) are forced on the right-hand side by the partial derivative of the error function $\mathcal{J}$ with respect to the reconstructed mean flow $\vpred$. This latter can be easily derived from Eq.~\ref{eq:cost_function} as
\begin{equation}
    \dfrac{\partial\mathcal{J}}{\partial\vpred} = 
    -\dfrac{\partial \mathcal{M}}{\partial \vpred}
    \left(\bar{\mathbf{m}} -\mathcal{M}(\vpred)\right).
    \label{eq:depsilon_du}
\end{equation}
Finally, the partial derivative of Eq.~\ref{eq:lagrangian_augmented} with respect to the forcing term $\ff$ yields:
\begin{equation}
    \dfrac{\partial\mathcal{J}}{\partial\ff} = \vpred ^\dag.
    \label{eq:depsilo_df}
\end{equation}
More details can be found in \cite{foures2014data}, where the case of localised measurements is also included in the analytical description of the direct-adjoint loop.
\medskip

By exploiting the gradients of the cost function $\mathcal{J}$ with respect to the control variable $\ff$ (Eq.~\ref{eq:depsilo_df}), a gradient descent algorithm can be applied to optimise $\ff$ and iteratively converge towards the optimal solution that minimises the cost function $\mathcal{J}$ (Eq.~\ref{eq:cost_function}). Specifically, the iterative direct-adjoint process refines the forcing term $\ff$, ensuring that the RANS model accurately captures the mean flow characteristics observed in high-fidelity DNS data. To summarise the process algorithmically, the adjoint optimisation procedure involves the following steps:
\begin{enumerate}
    \item \textbf{Initialisation} -- an initial guess for the control variable $\ff$ is chosen. We start from $\ff = 0$ to ensure the divergence-free property ($\nabla \cdot \ff = \mathbf{0}$) and consistency with the adopted boundary conditions \citep{foures2014data}. 
    \item \textbf{Forward step} -- a prediction of the mean flow $\vpred$ is obtained given the forcing term $\ff$, by solving Eq.~\ref{Eq:RANS}.
    \item \textbf{Cost function evaluation} -- the distance between the reconstructed mean flow $\vpred$ and the ground truth mean flow $\mean\vdir$ is assessed using the cost function $\mathcal{J}$ (Eq.~\ref{eq:cost_function}). 
    \item \textbf{Adjoint step} -- the adjoint equations (Eq.~\ref{eq:adjoint_NS}), forced by the term in Eq.~\ref{eq:depsilon_du}, are solved to find $\vpred^\dag$. The term in Eq.~\ref{eq:depsilo_df} expresses the variation of the cost function $\mathcal{J}$ with respect to the control variable $\ff$. 
    \item \textbf{Control variable update} -- using this gradient, corresponding to the adjoint variable $\vpred^\dag$, the forcing term $\ff$ is adjusted as:
    \begin{equation}
        \ff^{(n+1)} = 
        \ff^{(n)} +\beta\dfrac{\partial\mathcal{J}^{(n)}}{\partial \ff^{(n)}} =
        \ff^{(n)} +\beta\vpred^{\dag (n)},
    \end{equation}
    where the superscript $^{(n)}$ indicates the $n$-th iteration of the optimisation loop and $\beta$ a coefficient related to the step along the gradient direction. Note that the variable $\vpred^\dag$ is physically divergence-free, thus justifying the initialisation of $\ff$ as a null field. 
\end{enumerate}
The direct-adjoint process is iteratively repeated until the cost function $\mathcal{J}$ (Eq.~\ref{eq:cost_function}) drops below a predefined threshold, based on the required accuracy.

%% file: TexFiles/5_numerics.tex
\section{Numerical Setup}\label{sec:numerical}

\begin{figure}[t]
\centering
\includegraphics[width=.8\textwidth]{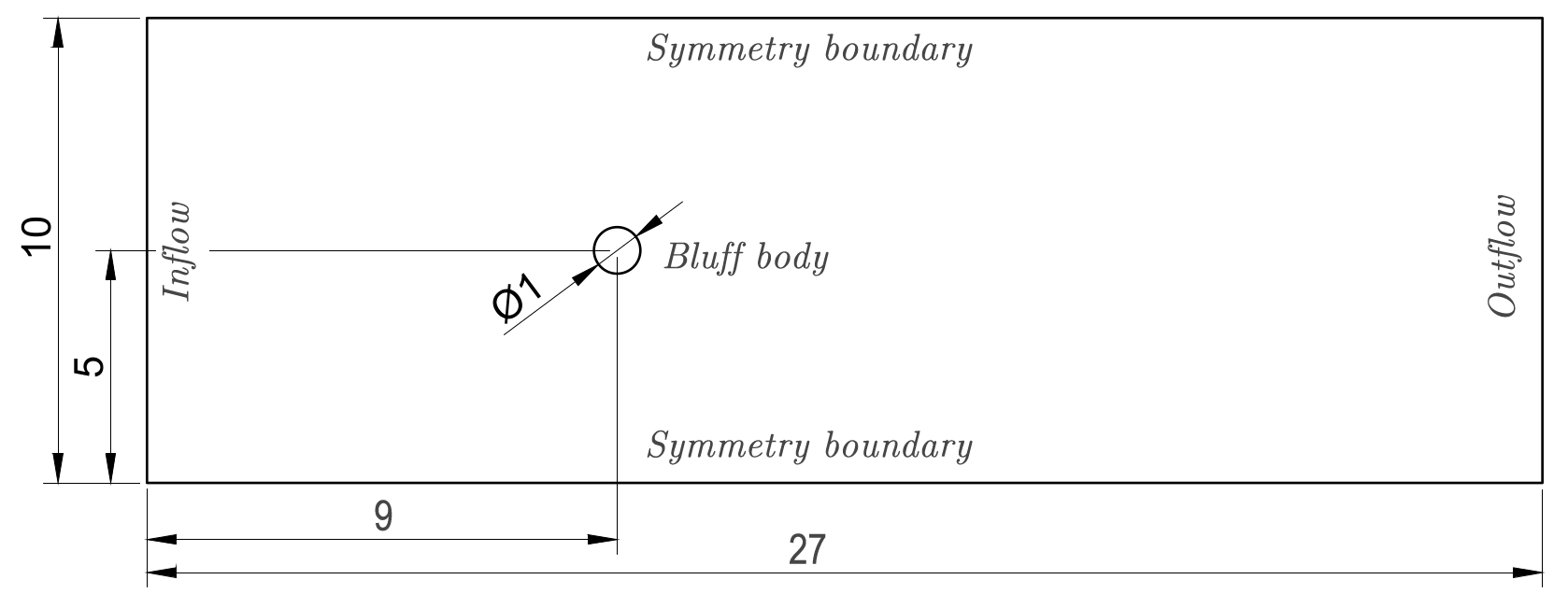}
\caption{
Sketch of the computational domain geometry, where the diameter of the circumscribed circle around the bluff body, along with the height and length of the domain, are provided in non-dimensional units}
\label{fig:computational_domain}
\end{figure}

The unsteady wake behind a bluff body is a well-established benchmark in CFD. As a reference case, the cylinder bluff body exhibits stable behaviour up to a critical Reynolds number, $Re_{c} \cong 46.7$ \citep{provansal1987benard, giannetti2007structural}. Beyond this threshold, irregular velocity fluctuations appear, accompanied by periodic vortex formation \citep{roshko1954drag}, and the unsteady flow evolves into a limit cycle known as the von Karman street. This phenomenon is observable up to $Re = 150$ for 2D cases \citep{roshko1954drag}. At these $Re$, dynamics can also settle in quasi-periodic regimes or show chaotic behaviour, see \emph{i.e.} the case of fluidic pinball \cite{deng2020low}. This study focuses on 2D scenarios exhibiting periodic or quasi-periodic behaviour within the range $50 \le Re \le 150$. In our numerical setup, the characteristic dimension is the diameter $D$ of the circumscribed circle around the bluff body. Based on this dimension, the computational domain extends $L_x = 27$ units in the streamwise direction and $L_y = 10$ units in the transverse direction. The system's origin $O(0, 0)$ is positioned $\Delta x = 9$ units downstream from the inlet and $\Delta y = 5$ units from the symmetry boundaries. A pictorial sketch of the geometric configuration of the computational domain is shown in Fig.~\ref{fig:computational_domain}.

The flow evolves from left to right with a dimensionless uniform velocity $\vdir = (1, 0)^T$, normalised by the reference velocity $U_{\infty}$ of the undisturbed flow. Boundary conditions follow the setup described by \cite{foures2014data}. For the direct NS equations (Eq.~\ref{eq:unsteadyNS}), they are as follows:
\begin{eqnarray}
\begin{cases}
\begin{array}{r l}
u = 1, \hspace{5pt} v = 0 &\text{at the inlet,} \\
u = 0, \hspace{5pt} v = 0 &\text{on the cylinder surface,}\\
\partial_y u = 0, \hspace{5pt} v = 0 &\text{on symmetry boundaries,}\\
\invRe\partial_x u - p = 0, \hspace{5pt} \partial_x v = 0 &\text{at the outlet}.\\
\end{array}
\end{cases}
\end{eqnarray}
For the adjoint NS equations (Eq.~\ref{eq:adjoint_NS}), the boundary conditions are:
\begin{eqnarray}
\begin{cases}
\begin{array}{r l}
u^\dag = 1, \hspace{5pt} v^\dag = 0 &\text{at the inlet,} \\
u^\dag = 0, \hspace{5pt} v^\dag = 0 &\text{on the cylinder surface,}\\
\partial_y u^\dag = 0, \hspace{5pt} v^\dag = 0 &\text{on symmetry boundaries,}\\
\invRe\partial_x u^\dag + p^\dag = -uu^\dag, \hspace{5pt} \invRe\partial_x v^\dag = -uv^\dag &\text{at the outlet}.\\
\end{array}
\end{cases}
\end{eqnarray}
Simulations start with null flow fields at $t=0$. Required statistics, such as the mean flow $\mean\vdir$ and forcing stress term $\ff$, are computed on-the-fly. The final simulation time $T$ is determined by a convergence criterion, specifically when the L2-norm difference between consecutive mean flows falls below $10^{-8}$.

Spatial discretisation is achieved using a Finite Element Method (FEM) approach via the \texttt{FEniCS} Python library \citep{alnaes2015fenics}, with time integration handled by a second-order Backward Differentiation Formula (BDF). Meshes are refined near the obstacle and in the wake region to accurately capture flow dynamics. Depending on the specific case, they typically consist of around $13,500$ nodes on average. Fig.~\ref{fig:mean_vort}$a$ depicts the streamwise component of the mean flow $\mean \vdir$ along with the vorticity isolines $\omega = \nabla \times \vdir$, while Fig.~\ref{fig:mean_vort}$b$ shows the streamwise component of the closure term $\ff$ for the cylinder bluff body reference case at $Re = 150$.

%% file: TexFiles/6_ML.tex
\section{Graph Neural Network overview}\label{sec:gnn_overview} 
In this section, we present an overview of the GNN architecture used in this study.
While a comprehensive description of the GNN  architecture can be found in Hamilton \emph{et al.} \cite{hamilton2020graph}, our focus here is to describe the specific implementation used in this work, tailored to the reconstruction of mean flow fields in CFD scenarios. In particular, we describe how unstructured mesh data are encoded as graphs, how physical quantities such as mean velocity fields and geometrical distances are integrated into the model as node and edge features respectively, and how the GNN outputs the predicted forcing term required to solve the RANS equations. The implementation relies on the \texttt{PyTorch Geometric} library \citep{fey2019fast}, which provides an efficient and scalable infrastructure for the Message Passing (MP) algorithms at the core of the GNN model.\\
Sec.~\ref{subsec:mp_algorithm} delves into the MP process adopted in our GNN, Sec.~\ref{subsec:data_structure} outlines the specific structure of the input data used to encode CFD problems as graphs; the GNN training and inference algorithm is presented in Sec.~\ref{subsec:training_process_algorithm}.
For readers already familiar with Quattromini \emph{et al.} \cite{quattromini2023operator}, we note that the GNN architecture described here is unchanged. However, it is employed in this work with a fundamentally different objective; those readers may wish to skip directly to Section~\ref{sec:methodology}, where the novel data assimilation strategy is introduced.

\subsection{Message Passing Process}\label{subsec:mp_algorithm}
In GNNs, nodes iteratively exchange information with their neighbours to update their latent representations. This process, known as MP, allows GNNs to capture complex dependencies and patterns within data by considering the edges as important information relevant to the problem. Depending on the graph's size, the MP process can be repeated an arbitrary number of times, defined by a hyperparameter $k$. A detailed list of the hyperparameters used is provided in App.~\ref{appendixB:hyperparameter}. The MP process is node-centered and consists of three fundamental steps:
\begin{enumerate}
    \item \textbf{Message Creation}: Each node $i$ begins with an embedding state represented by a vector $\mathbf{h}_i$. Initially set to zero, this vector accumulates and processes information during the MP iterations. The dimension $d_h$ of $\mathbf{h}_i$ is constant across all nodes and is a key model hyperparameter. This dimension determines the expressivity of the GNN, which reflects its ability to model complex functions \citep{guhring2020expressivity, hornik1989multilayer}. Note that the embedded state itself does not have a direct physical interpretation, 
    while the distance matrix indeed captures some
physical relation between the nodes. \medskip
    
    \item \textbf{Message Propagation}: Information is propagated between nodes. To capture the convective and diffusive dynamics of the underlying CFD system, messages are exchanged bidirectionally between connected nodes. For a pair of connected nodes $i$ and $j$, with a directed connection $\mathbf{a}_{ij}$ from $i$ to $j$, the message generated is defined as:
    \begin{equation}
    \boldsymbol{\phi}_{i,j}^{(k)} = \zeta^{(k)}(\mathbf{h}_i^{(k-1)}, \mathbf{a}_{ij}, \mathbf{h}_j^{(k-1)}),
    \label{eq:phi}
    \end{equation}
    where $\mathbf{h}_i^{(k-1)}$ is the embedded state from the previous MP layer $(k-1)$, and $\zeta^{(k)}$ is a differentiable operator, such as a Multi-Layer Perceptron (MLP) \citep{goodfellow2016deep}. By swapping the indices $i$ and $j$ in Eq.~\ref{eq:phi}, we obtain the message flowing from $j$ to $i$. Depending on the number of $j$ connected nodes in the neighbouring set of $i$, namely $\mathcal{N}_i$, for each node $i$ the global outgoing message is then computed as:    
    \begin{equation}
        \boldsymbol{\phi}_{i,\rightarrow} = \bigoplus_{j\in\mathcal{N}_i} \boldsymbol{\phi}_{i,j}
    \end{equation}
    where $\bigoplus$ represents a differentiable, permutation-invariant function (\emph{e.g.}, sum, mean, or max). In our study, we choose the mean function to preserve permutation invariance, which is crucial when working with unstructured meshes where the neighbourhood size of each node may vary. This ensures that the model can generalise across different mesh configurations.\medskip
    
    \item \textbf{Message Aggregation}: Each node $i$ aggregates the received information to update its embedded state $\mathbf{h}_i^{(k)}$:
    \begin{equation}
    \mathbf{h}_i^{(k)} = \mathbf{h}_i^{(k-1)} + \alpha \Psi^{(k)}(\mathbf{h}_i^{(k-1)}, \mathbf{G}_i, \boldsymbol{\phi}_{i,\rightarrow}^{(k)}, \boldsymbol{\phi}_{i,\leftarrow}^{(k)}, \boldsymbol{\phi}_{i,\circlearrowright}^{(k)}),
    \end{equation}
    where $\mathbf{G}_i = \{\overline{\mathbf{u}}, Re\}$ represents external quantities injected into the GNN (mean flow $\mean\vdir$ and Reynolds number $Re$ in our case), provided at each update $k$. The terms $\boldsymbol{\phi}_{i,\rightarrow}^{(k)}$ and $\boldsymbol{\phi}_{i,\leftarrow}^{(k)}$ represent respectively the message sent to and received from all the neighbouring nodes. The term $\boldsymbol{\phi}_{i,\circlearrowright}^{(k)}$ is the self-message that the node $i$ send to itself in order to maintain the node's own information while aggregating messages from its neighbours. Their mathematical definition, with the appropriate change in notation, is expressed in Eq.~\ref{eq:phi}. The term $\Psi^{(k)}$ is a differentiable operator, typically an MLP, used to handle together the gathered information. The term $\alpha$ is a hyperparameter relaxation coefficient controlling the update scale (App.~\ref{appendixB:hyperparameter}).
\end{enumerate}

At the end of the MP process, each node's embedded state has been updated $k$ times, incorporating information from its neighbours. The choice of $k$ is crucial for the model's generalisation across graphs of varying sizes. As shown by \cite{nastorg2024scalable}, $k$ should be proportional to the graph's diameter to ensure effective information propagation \citep{donon2020deep}. However, since the graphs in the dataset used in this study exhibit a relatively consistent diameter, we opted to optimise this hyperparameter using genetic algorithms (App.~\ref{appendixB:hyperparameter}). Additionally, \cite{nastorg2024scalable} explored recurrent or adaptive architectures, where $k$ could dynamically adjust based on the graph's structure. Once the MP process concludes, the latest $k$-updated embedded state of each node is projected back into the physical space to predict the desired target, in this case, the forcing stress term $\ff$. A differentiable operator such as an MLP (decoder $D$) handles this operation.

\begin{figure}[t]
\centering
\includegraphics[width=.825\textwidth]{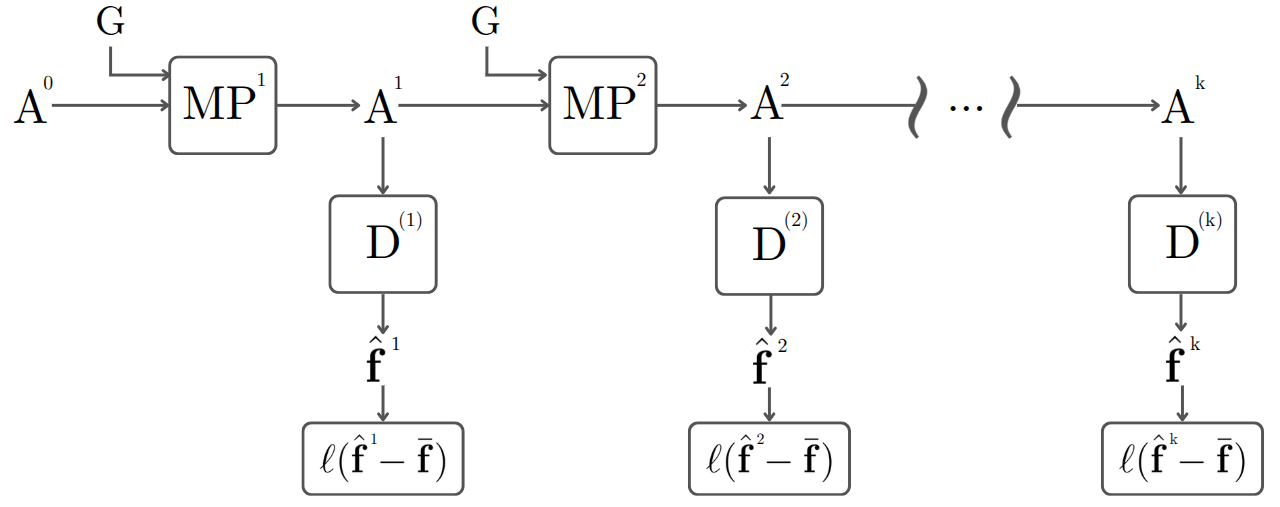}
\caption{
The overall framework of our GNN training process. ${MP}^k$ denotes the message passing algorithms, ${D}^{(k)}$ are the $k$ decoders, which are trainable MLPs, $\textbf{A}^k$ are the $k$ matrices containing the embedded states of each node, and $\textbf{G}$ is the vector containing the input injected into the GNN. The figure is inspired by \cite{donon2020deep}}
\label{fig:gnn_training}
\end{figure}

\subsection{Data Structuring}\label{subsec:data_structure}
Applying GNNs to unstructured CFD data requires representing the CFD mesh as a graph. In order to obtain the CFD-GNN interface, we align each mesh node with a GNN node. To this end, we structure the data into tensors that maintain the mesh's adjacency properties. Specifically, for each case in the ground truth dataset, we construct:
\begin{itemize}
    \item A matrix $\mathbf{A} \in \mathbb{R}^{n_i \times d_h}$, 
    \[
\textbf{A}_{n_i,d_h} =
\begin{bmatrix}
a_{1,1} & a_{1,2} & \cdots & a_{1,d_h} \\
a_{2,1} & a_{2,2} & \cdots & a_{2,d_h} \\
\vdots  & \vdots  & \ddots & \vdots  \\
a_{n_i,1} & a_{n_i,2} & \cdots & a_{n_i,d_h}
\end{bmatrix},
\]  
    where $n_i$ is the number of nodes in the mesh and $d_h$ is the dimension of the embedded state. This matrix $\mathbf{A}$ stacks together all the embedded arrays $h_i$ defined on all the nodes.
    \item A matrix $\mathbf{C} \in \mathbb{R}^{c \times 2}$, where $c$ is the number of edges in the mesh, defining the node connections. This matrix represents the adjacency scheme of the mesh. 
    \item A matrix $\mathbf{E} \in \mathbb{R}^{c \times 2}$, containing the distances between connected nodes in the $x$ and $y$ directions. This matrix captures the geometric properties of the mesh's adjacency scheme, expressed as the node distances.
\[
\textbf{C}_{c,2} =
\begin{bmatrix}
i_1 & j_1\\
i_2 & j_2\\
\vdots  & \vdots \\
i_c & j_c
\end{bmatrix},
\qquad
\textbf{E}_{c,2} =
\begin{bmatrix}
x_{i_1} - x_{j_1} & y_{i_1} - y_{j_1}\\
x_{i_2} - x_{j_2} & y_{i_2} - y_{j_2}\\
\vdots  & \vdots \\
x_{i_c} - x_{j_c} & y_{i_c} - y_{j_c}\\
\end{bmatrix}.
\]
\end{itemize}
Each column of $\mathbf{A}$ serves as a feature vector for neurons in the MLPs used in the GNN ($\zeta$, $\Psi$, and the decoder $D$). The MLPs architecture is defined by the dimension $d_h$ of the embedded state, while the number of nodes $n_i$ corresponds to the feature count per neuron. This approach enables the use of the same MLP architectures across different CFD simulations, regardless of geometry or node count, as the number of nodes does not affect the underlying structure of the MLP. This approach makes the GNNs particularly well-suited for interacting with unstructured meshes and learning from various geometries and configurations.

\subsection{GNN Training Algorithm}\label{subsec:training_process_algorithm}
The GNN training framework is shown in Fig.~\ref{fig:gnn_training}. The process begins with $\mathbf{A}^{0}$, a matrix of zero-initialized embedded states (see Sec.~\ref{subsec:data_structure}). This matrix, along with external inputs $\mathbf{G}$ (mean flow $\mean\vdir$ and Reynolds number $Re$), is fed into the first message passing layer $\text{MP}^{1}$. The updated embedded state matrix $\mathbf{A}^{1}$ is then passed through the decoder $D^{1}$, an MLP responsible for reconstructing the physical state $\hat{\ff}^{1}$. The predicted forcing term $\hat{\ff}^{1}$ is compared to the DNS ground truth $\ff$ using a loss function: 
\begin{equation}
    \ell^k = \frac{1}{n_i}\sum_{i=1}^{n_i}(\ff^k_i - \hat{\ff_i})^2,
    \label{eq:loss_function}
\end{equation}
where $n_i$ is the number of nodes, and $k$ is the layer number in the MP process. This procedure is repeated across the $k$ layers of the GNN. Following \cite{donon2020deep}, the intermediate losses from all layers of the MP process are combined in a global loss function $L$ to robustify the learning process:
\begin{equation}
    L = \sum_{k=1}^{\bar{k}} \gamma^{\bar{k}-k} \cdot \ell^k,
    \label{eq:global_cost_generic_GNN}
\end{equation}
where $\bar{k}$ is the number of layers, and $\gamma$ is a hyperparameter controlling the weight of each loss term (see App.~\ref{appendixB:hyperparameter}). As the MP process goes on, each node collects more and more information. This exponential weighting ensures that later updates, which are richer in information, have greater influence on the learning process.

%% file: TexFiles/7_methodology.tex
\section{Methodology}\label{sec:methodology}

\begin{figure}[t]
\centering
\includegraphics[width=1\textwidth]{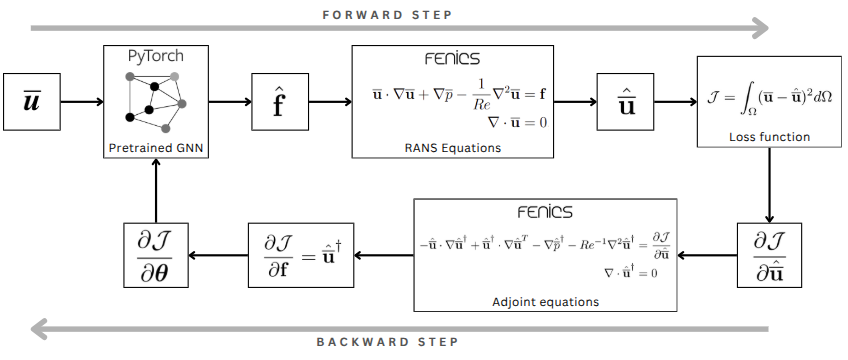}
\caption{
End-to-end training loop: $\mean\vdir$ is the GNN's input mean flow, $\fpred$ is the GNN's predicted forcing stress term, $\boldsymbol{\theta}$ represents the GNN's trainable parameters, and $\mathcal{J}(\hat{\mean\vdir})$ is the cost function to minimise. For simplicity of notation, we consider the case where the ground truth corresponds to the entire flow field.}\label{fig:GNN_CFD_scheme}
\end{figure}

This section describes the data assimilation scheme developed in this study, which combines the training process of a neural model based on GNN (Sec.~\ref{sec:gnn_overview}) with the RANS equations (Sec.~\ref{sec:governing_equations}). The approach primarily focuses on using analytically derived gradients through the adjoint method (Sec.~\ref{subsec:adjoint_methods}) to enhance the GNN learning process and ensure physical consistency in its predictions. This scheme will be referred to as the Physics-Constrained Graph Neural Network (PhyCo-GNN). The complete end-to-end training process is illustrated in Fig.~\ref{fig:GNN_CFD_scheme}, with technical details provided in Sec.~\ref{subsec:training_process} regarding the training process, in Sec.~\ref{subsec:pre-training} for the GNN pre-training phase, and in Sec.~\ref{subsec:double_loss_function} for the approach used to handle the transition from pre-training to the main training phase of the GNN.

\subsection{Algorithm and training process}\label{subsec:training_process}
With reference to Fig.~\ref{fig:GNN_CFD_scheme}, the global training process can be ideally divided into two phases: the forward step and the backward step.

The forward step begins by associating an embedded state vector $\mathbf{h}_0$, initially set to zero, as a node feature to each node of the GNN. These vectors, one for each mesh node, define the initial latent representation of the graph. Each node of the graph corresponds to a point in the computational mesh, and this mapping is preserved through the data structure described in Sec.~\ref{subsec:data_structure}.
The Euclidean distances between nodes are used as edge features to define the graph connectivity and encoding the local geometric relationships. The constructed graph is passed through a pre-trained GNN (Sec.~\ref{subsec:pre-training}) and the state vector on each node is updated through a $k$-step message passing process, as outlined in Sec.~\ref{subsec:mp_algorithm}. During each message passing iteration, nodes exchange and aggregate information with their neighbours leveraging the geometric edge features. External physical quantities (the mean flow field $\mean{\vdir}$ and the Reynolds number $Re$), are injected during the aggregation step at each layer, providing the GNN with the necessary context to learn physically meaningful patterns. After $k$ message passing iterations, the GNN outputs a predicted forcing term $\fpred$, which represents an estimate of the Reynolds stress closure term.
This predicted forcing term $\fpred$ is used as the closure term in the RANS equations (Eq.~\ref{Eq:RANS}). These equations are numerically solved in a FEM framework using the Python library \verb|FEniCS| \citep{alnaes2015fenics}, producing a reconstructed mean flow field $\vpred$. This result is then compared with the mean flow ground truth $\mean\vdir$, obtained from the DNS, to compute the loss function $\mathcal{J}$, as expressed in Eq.~\ref{eq:cost_function}. \\

The second phase, the backward step, starts with the requirement to compute the derivative of the loss function $\mathcal{J}$ with respect to the trainable $\boldsymbol{\theta}$ parameters of the GNN. The gradient chain rule for this term can be mathematically expressed as:
\begin{equation}
\frac{\partial \mathcal{J}}{\partial\boldsymbol{\theta}} = \frac{\partial \mathcal{J}}{\partial\vpred} \cdot \frac{\partial\vpred}{\partial\fpred} \cdot \frac{\partial\fpred}{\partial\boldsymbol{\theta}} = \frac{\partial \mathcal{J}}{\partial\fpred} \cdot \frac{\partial\fpred}{\partial\boldsymbol{\theta}}.\label{eq:chain_rule}
\end{equation}
The first term, $\frac{\partial \mathcal{J}}{\partial \fpred}$, on the right-hand side is obtained from Eq.~\ref{eq:depsilo_df} after solving the adjoint equations (Eq.~\ref{eq:adjoint_NS}) with the support of Eq.~\ref{eq:depsilon_du}. This yields an analytical gradient of the cost function $\mathcal{J}$ with respect to $\fpred$.
The second term, $\frac{\partial \fpred}{\partial \boldsymbol{\theta}}$, represents the gradient of the GNN's output with respect to its $\boldsymbol{\theta}$ trainable parameters, which can be computed efficiently using the \texttt{PyTorch Geometric} automatic differentiation.\\ 
These two gradients -- one analytically derived and discretised using \verb|FEniCS|, and the other numerically computed via automatic differentiation in \texttt{PyTorch Geometric} \citep{fey2019fast} -- are combined to complete the chain rule (Eq.~\ref{eq:chain_rule}) and obtain the full derivative that is used to update the GNN learnable parameters via gradient descent. The full forward-backward process is repeated iteratively until convergence, i.e., when the cost function $\mathcal{J}$ falls below a predefined threshold.

\subsection{On the pre-training step}\label{subsec:pre-training}
A crucial step in the algorithm is the GNN model's pre-training phase. We can motivate this choice from two different perspectives. From a theoretical Bayesian viewpoint, the objective is to infer the posterior distribution of the GNN parameters $\theta$ given observations of the mean velocity field $\bar{\mathbf{u}}$, i.e., $p(\theta \mid \bar{\mathbf{u}})$. Since the forcing field $\ff$ is not observed directly, but acts as a latent variable linking $\theta$ to $\bar{\mathbf{u}}$ via the RANS equations, we introduce $\mathbf{f}$ as an auxiliary variable and marginalise over it. This yields:
\begin{equation}
p(\theta \mid \bar{\mathbf{u}}) \propto \int p(\bar{\mathbf{u}} \mid \mathbf{f}) \, p(\mathbf{f} \mid \theta) \, p(\theta) \, d\mathbf{f},
\end{equation}
where $p(\bar{\mathbf{u}} \mid \mathbf{f})$ models the likelihood of the observed mean flow given a candidate forcing field; $p(\mathbf{f} \mid \theta)$ is the GNN model predicting $\mathbf{f}$ from input features and $p(\theta)$ is the prior over the model parameters, which we assume to be uniform in the absence of other information. This decomposition highlights the role of $\mathbf{f}$ as a latent intermediate quantity through which the model learns to match the observed $\bar{\mathbf{u}}$. Pretraining the GNN on DNS-based samples of $\mathbf{f}$ provides a meaningful initialisation for $p(\mathbf{f} \mid \theta)$, thus guiding the posterior inference on $\theta$ in a more stable and efficient way during the later reconstruction of $\bar{\mathbf{u}}$. Moreover, the GNN's weights and biases are set using a default initialisation \citep{he2015delving}, meaning that early predictions from the GNN are non-physical and cannot reliably be used in the forward step, where the forcing term is introduced into the RANS equations to compute the mean flow (Sec.~\ref{subsec:training_process}). In fact, the solution to the RANS problem may not exist if the initial guess for the forcing term, $\fpred$, is too far from a physically meaningful value. Thus, from the numerical viewpoint, the pre-training phase stabilises the GNN's output and mitigates this issue, making the prediction of the forcing stress term $\fpred$ suitable for integration into the RANS equations.\\

The pre-trained model is obtained through pure supervised learning \citep{quattromini2023operator}, where the mean flow $\mean\vdir$ (and Reynolds number $Re$) are used as input, and the forcing stress term $\ff$ from DNS data is the target. The loss function used in this phase is the Mean Squared Error (MSE) loss $\mathcal{L}$, already discussed in Sec.~\ref{subsec:training_process_algorithm}.
where $n_i$ is the number of nodes in the GNN. 
The number of epochs required to reach the desired closure term accuracy is not fixed a priori, but is determined based on the convergence behaviour of the GNN during the pre-training phase. In particular, the transition to the main training phase occurs only when the GNN has learned to produce physically plausible forcing terms $\fpred$, which allow the RANS equations to be solved via the FEM framework. 
To ensure consistency across test cases, a standard protocol is adopted: for simple reference scenarios, such as those presented in Section~\ref{subsec:baseline}, the pre-training is fixed at 500 epochs. For more complex cases, which involve larger datasets, the pre-training phase is extended to 1000 epochs. This choice reflects the increased training effort required to reach comparable levels of accuracy in more challenging settings. \\

In this context, the closure term accuracy refers to the level of precision necessary for the GNN to generate predictions that allow the FEM solver to successfully solve the RANS equations. Throughout the pre-training phase, the GNN's predictions are periodically evaluated by solving a test FEM step. If the solver converges and accurately resolves the RANS equations using the GNN-predicted closure term, the pre-training phase is considered complete. This ensures that the GNN has learned a reliable and accurate representation of the closure term, making it suitable for the full training process.

\subsection{On the loss function}\label{subsec:double_loss_function}
During the pre-training step (Sec.~\ref{subsec:pre-training}), the GNN model is updated using a loss function designed to align the model's predictions with the available DNS data. As previously mentioned, this phase serves as a warm-up for the subsequent main training (Sec.~\ref{subsec:training_process}). However, when pre-training ends and the main training begins, a different loss function is adopted, as can be seen by comparing Eq.~\ref{eq:global_cost_generic_GNN} with Eq.~\ref{eq:cost_function}. This change may negatively affect convergence and destabilise the training process, as the two optimisation landscapes can be quite different. To mitigate this risk, both loss functions are retained during the main training phase and combined using a weight coefficient, $\beta$. In particular, the loss function $\mathcal{L}$ (Eq.~\ref{eq:global_cost_generic_GNN}), associated with supervised pre-training, continues to enforce a data-driven alignment and ensures "continuity" in the optimisation process. On the other hand, the term $\mathcal{J}$ (Eq.~\ref{eq:cost_function}), corresponding to the physics-constrained loss, minimises the mean flow reconstruction error. Thus, the global loss function for the main training loop (Sec.~\ref{subsec:training_process}) evolves as:
\begin{equation}
    \mathcal{H} = (1-\beta)\mathcal{L} + \beta\mathcal{J} = (1-\beta)\left(\frac{1}{n} \sum_{i=1}^{n} (\ff_i - \fpred_i)^2\right) + \beta \left ( \frac{1}{2}\int_\Omega \left(\bar{\mathbf{m}} - \mathcal{M}(\hat{\mean\vdir})\right)^2 d\Omega \right ).
    \label{eq:loss_global}
\end{equation}
This strategy ensures a smooth transition between the two optimisation steps by adjusting the relative importance of the pre-training and main training loss functions.

The next section discusses the results, showing that the converged GNN model effectively predicts a forcing term $\hat{\ff}$ that aligns with the ground truth, $\ff$ allowing an effective learned model to reconstruct the mean flow $\mean\vdir$.

%% file: TexFiles/8_results.tex
\section{Results}\label{sec:results}

\begin{figure}[t]
    \centering
    \includegraphics[width=0.99\textwidth]{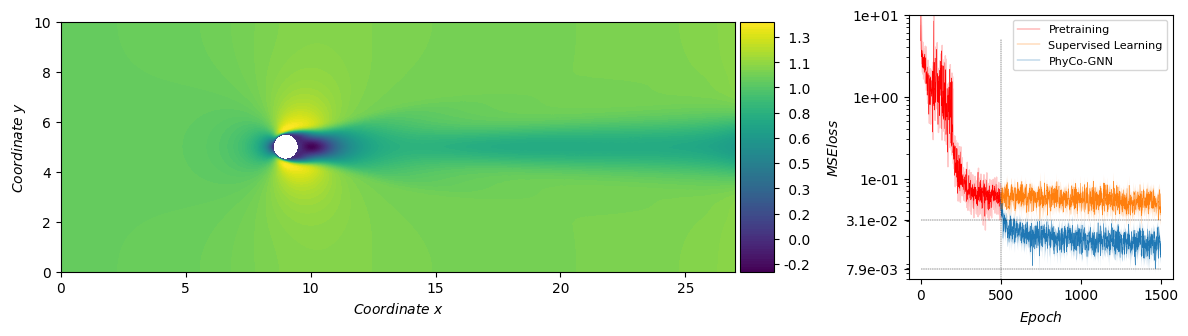}
    \includegraphics[width=0.99\textwidth]{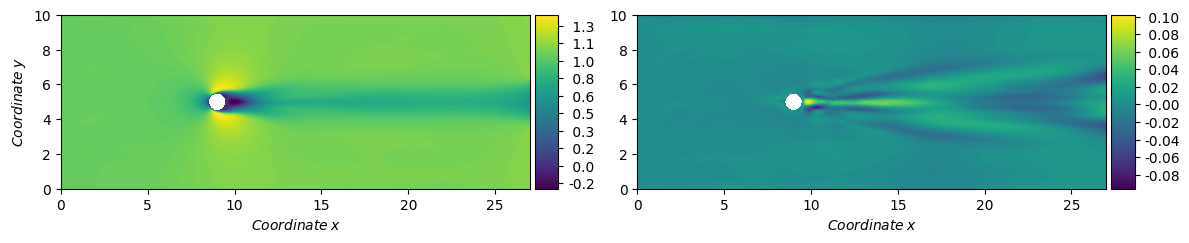}
    
    \begin{picture}(0,0)
        \put(-200, 216){$a)$}
        \put(  90, 216){$b)$} 
        \put(-200, 87){$c)$}
        \put(  25, 87){$d)$} 
    \end{picture}
    \vspace{-.5cm}
    \caption{Training Dataset: $1$ mean flow-forcing pair ($Re = 150$); the GNN input of the mean flow from the ground truth is shown in ($a$); ($b$) Loss curves for the pure supervised approach (orange line), the proposed PhyCo-GNN method (blue line), and the pre-training phase (red line). Shadow colors highlight standard deviations computed over 5 independent training runs with different parameter initialisations. The two horizontal dotted lines indicate the minimum values of the supervised and proposed methods. ($c$) Reconstructed mean flow obtained with the PhyCo-GNN approach. ($d$) Contour plot of the reconstruction error difference between the pure supervised approach and PhyCo-GNN} \label{fig:baseline_150}
\end{figure}

In this section, we present the improvements achieved using the PhyCo-GNN data assimilation scheme for reconstructing the mean flow field $\mean\vdir$. The tests are conducted across several scenarios, with a particular focus on reconstructing the mean flow from sparse measurements, noisy probes, and incomplete flow fields (inpainting). The models are compared with the pure supervised learning method, which serves as a baseline. The supervised learning approach is thoroughly discussed in \cite{quattromini2023operator}: the GNN model is trained solely by learning the forcing stress from the DNS data. Therefore, the cost function is chosen to minimise the discrepancy between the predicted forcing stress and the ground truth, without incorporating any constraints based on the system's physics. The mean flow $\mean\vdir$ is then computed by using the forcing stress modelled with the GNN as input to the RANS equations in Eq.~\ref{Eq:RANS}. In contrast, the hybrid data assimilation scheme introduced in this work incorporates a physics-based constraint during the training process of the GNN. To compare the two methods, we evaluate their training curves (after the pre-training phase) by identifying the minimum loss values achieved by each model during training. Unless otherwise stated, all test cases use the full velocity field from DNS as input; the sparse probe setting applies exclusively to the test in Sec.~\ref{subsec:sparse_measurements}. As a metric for comparison, we consider the percentage improvement, defined as:
\begin{equation}
    \mathcal{I}(\%) = \dfrac{min( \mathcal{J}_{\text{S}}) - min( \mathcal{J}_{\text{PC}})}{min( \mathcal{J}_{\text{S}})} \cdot 10^2,  
    \label{eq:percentage_improvement}
\end{equation}
where $min( \mathcal{J}_{\text{S}})$ and $min( \mathcal{J}_{\text{PC}})$ represent the minimum values of the loss function for mean flow reconstruction (Eq.~\ref{eq:cost_function}) in the supervised learning method and the adjoint-based PhyCo-GNN method, respectively. In the following, we introduce different learning tasks, focusing on the technical features of the method and discussing the improvements achieved in terms of mean flow reconstruction.

\subsection{Reference cases}\label{subsec:baseline}

\begin{figure}[ht]
    \centering
    \includegraphics[width=0.99\textwidth]{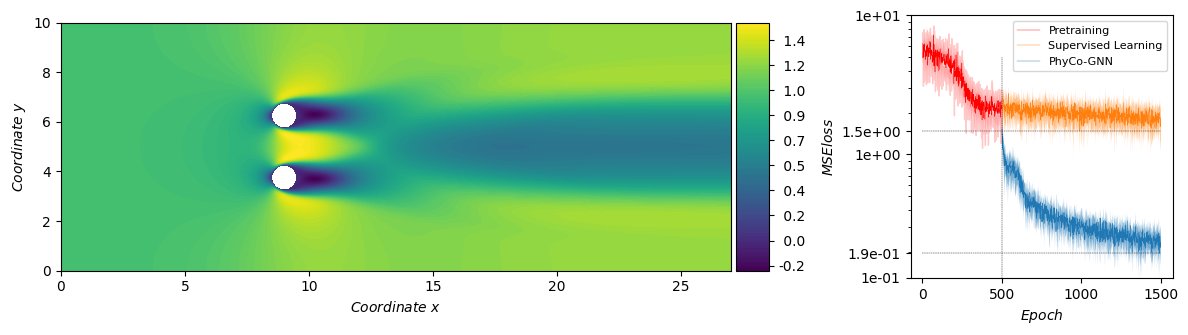}
    \includegraphics[width=0.99\textwidth]{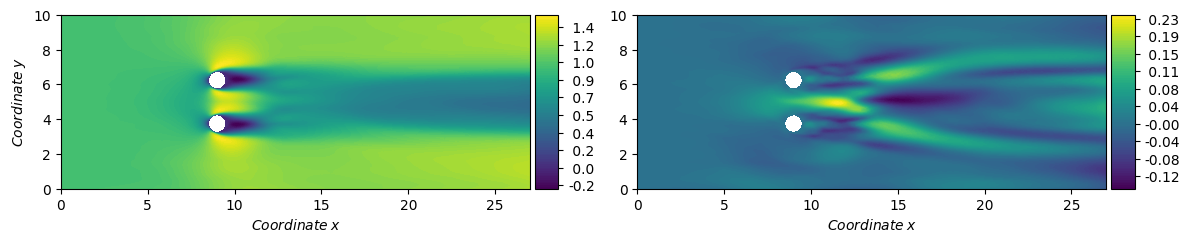}
    
    \begin{picture}(0,0)
        \put(-200, 216){$a)$}
        \put(  90, 216){$b)$} 
        \put(-200, 87){$c)$}
        \put(  25, 87){$d)$} 
    \end{picture}
    \vspace{-.5cm}
    \caption{Training Dataset: $1$ mean flow-forcing pair at $Re = 90$; the mean flow is shown in ($a$) and is used as GNN input; legend for ($b$), ($c$), and ($d$) as in Fig.~\ref{fig:baseline_150}}
    \label{fig:baseline_90_4}
\end{figure}

The first test case we consider is the reconstruction of the flow field when the input to the GNN is the complete mean flow $\mean\vdir$ and Reynolds number $Re$ over the entire computational domain $\Omega$. This test case is introduced as a reference for the method. Throughout this section, the mapping operator $\mathcal{M}$ used in the definition of the cost function is taken as the identity operator, implying that the comparison is performed over the full velocity field without any spatial filtering or subsampling.\\ We consider two cases of increasing complexity. The first case involves flow past a 2D cylinder at a Reynolds number of $Re = 150$. This case is well-documented in the literature, and its mean flow is shown in Fig.~\ref{fig:baseline_150}$a$. The training dataset includes only the mean flow $\mean\vdir$ (used as input to the GNN) and its corresponding forcing term $\ff$ (the GNN target). The training curves in Fig.~\ref{fig:baseline_150}$b$ show that, starting from the pre-training phase, the implementation of the approach described in this paper leads to a substantial improvement in mean flow reconstruction. Specifically, the improvement reaches a value of $\mathcal{I}=58.59\%$.

The second case involves a two side-by-side cylinder configuration, also known in the literature as the 'flip-flop' case, at a Reynolds number of $Re = 90$ \cite{carini2014origin}. Its mean flow is shown in Fig.~\ref{fig:baseline_90_4}$a$. The training curves for this case in Fig.~\ref{fig:baseline_90_4}$b$ demonstrate an even more pronounced improvement, with a reduction of $\mathcal{I}=82.90\%$ in the loss curve. The results not only indicate the broad adaptability of the PhyCo-GNN but also show how, in more complex scenarios, the underlying physics and governing equations play a crucial role in further enhancing the accuracy of the GNN model's predictions.

\subsection{Generalisation}\label{subsec:generalization}
In this section, we test the generalisation capabilities of the model obtained using the PhyCo-GNN scheme. The training dataset consists of three cases involving a 2D cylinder at Reynolds numbers of $Re = [90, 110, 130]$. In contrast, the validation dataset includes data points not present in the training set, corresponding to simulations of the flow around a 2D cylinder at Reynolds number $Re = 120$ for interpolation testing, and at $Re = 150$ for testing the extrapolation capabilities.
Throughout this generalisation test, the mapping operator $\mathcal{M}$ used in the cost function remains the identity operator.
\begin{figure}[ht]
    \centering
    \includegraphics[width=0.99\textwidth]{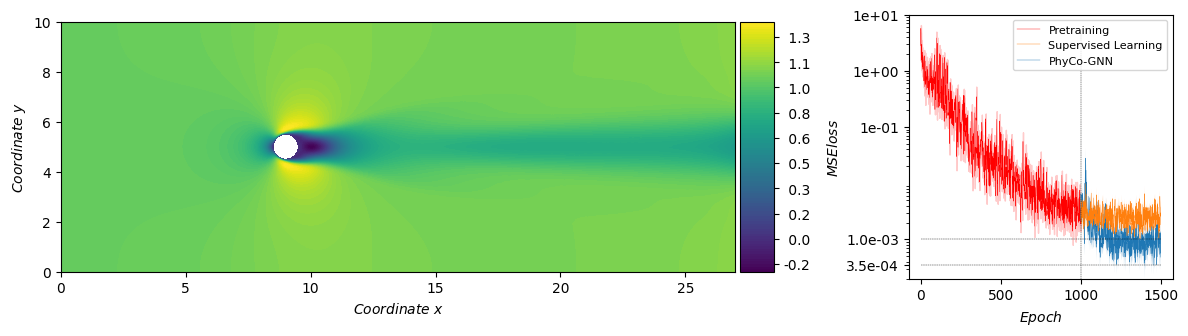}
    \includegraphics[width=0.99\textwidth]{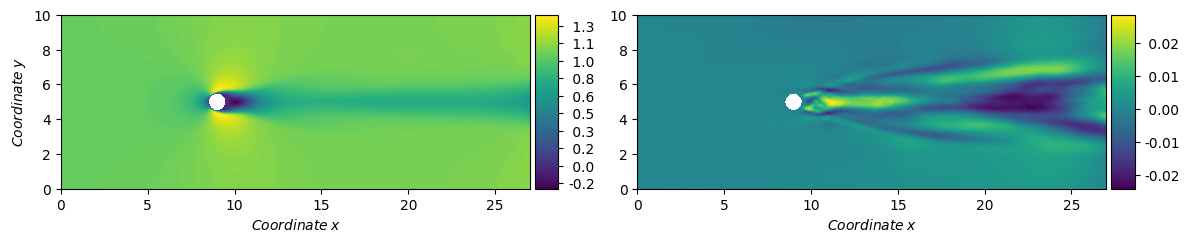}

    \begin{picture}(0,0)
        \put(-200, 216){$a)$}
        \put(  90, 216){$b)$} 
        \put(-200, 87){$c)$}
        \put(  25, 87){$d)$} 
    \end{picture}

    \vspace{-.5cm}
    \caption{Generalisation test -- training dataset: $3$ meanflow-forcing pairs at $Re = [90, 110, 130]$; Validation Dataset: $2$ meanflow-forcing pairs at $Re = [120, 150]$; ($a$) ground truth mean flow at $Re = 120$ from the validation dataset, used as GNN input; legend for ($b$), ($c$), and ($d$) as in Fig.~\ref{fig:baseline_150}}
    \label{fig:generaliz_ground_curves}
\end{figure}
In Fig.~\ref{fig:generaliz_ground_curves}$a$, the ground truth mean flow $\mean\vdir$ for the $Re = 120$ case is shown. Based on the validation cases, we observe an improvement in the mean flow reconstruction, with an average improvement of $\mathcal{I}=73.27\%$ across the entire validation dataset. Specifically, we achieve an improvement of $\mathcal{I}=78.96\%$ for the interpolation case at $Re = 120$, and $\mathcal{I}=13.96\%$ for the extrapolation case at $Re = 150$. The improvement on the training cases is $\mathcal{I}=40.16\%$ on average across the entire training dataset.

The primary objective of this test case is to demonstrate that the presented approach enhances the generalisation capabilities of the GNN model. To ensure clarity in our analysis, this generalisation test case is intentionally separated from the others. This separation allows for a focused evaluation of each individual test case, targeting the specific goals of those tests without introducing confounding variables related to generalisation. It should be noted that a more extensive assessment of the GNN’s ability to generalise across a range of geometric configurations -- such as variations in bluff body shapes or positions -- is beyond the scope of this study. These aspects have been examined in detail by Quattromini \emph{et al.} \cite{quattromini2023operator}, in combination with active learning strategies. In the present work, we focus on a simplified setting with limited generalisation in order to clearly evaluate the role of the physics-constrained training scheme in predicting the mean flow.

\subsection{Sparse Measurements}\label{subsec:sparse_measurements}
The learning task presented here involves the reconstruction of the mean flow over the entire computational domain using measurements from randomly distributed probes as input for the GNN. In this case, the mapping operator $\mathcal{M}$ appearing in the cost function is defined as the projection onto the sparse set of probe locations, so that the comparison between predicted and reference fields is performed only at the measurement points. The numerical treatment of the term forcing the adjoint equation -- based on the sparse measurements -- follows the one included in \cite{foures2014data}.
\begin{figure}
    \centering
    \includegraphics[width=0.99\textwidth]{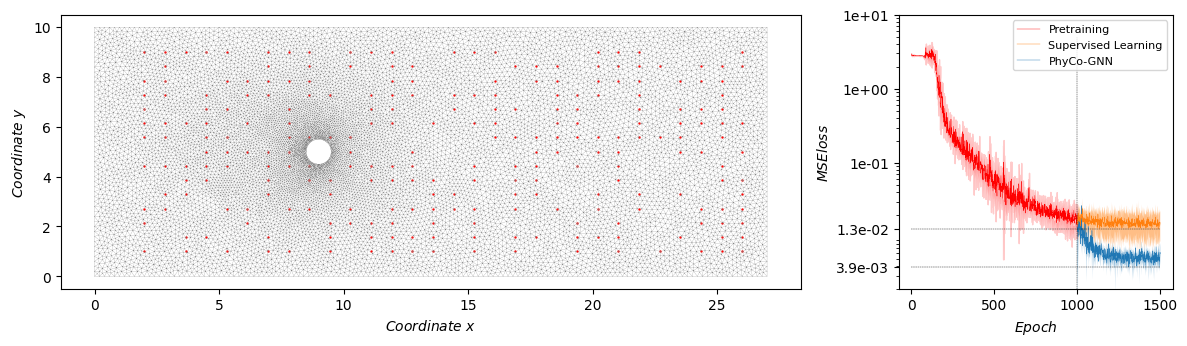}
    \includegraphics[width=0.99\textwidth]{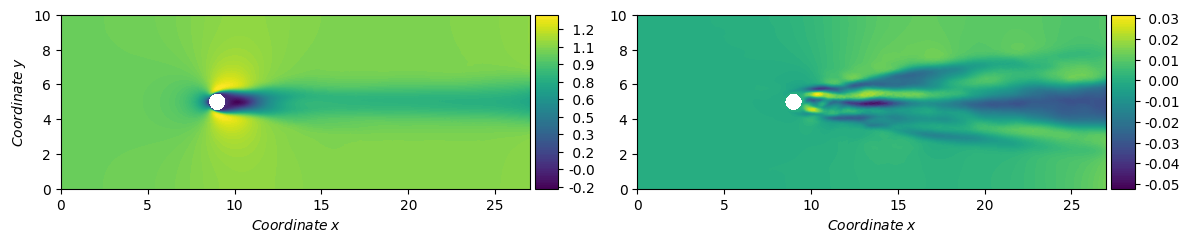}
    
    \begin{picture}(0,0)
        \put(-200, 216){$a)$}
        \put(  90, 216){$b)$} 
        \put(-200, 87){$c)$}
        \put(  25, 87){$d)$} 
    \end{picture}
    \vspace{-.5cm}
    \caption{Sparse measurements -- training dataset: 250 randomly distributed probes on $6$ mean flow-forcing pairs at $Re = [90, 110, 130]$, with two instances for each $Re$); ($a$) Random probes positioning on the mean flow; legend for ($b$), ($c$), and ($d$) as in Fig.~\ref{fig:baseline_150}}\label{fig:sparse_ground_curves}
\end{figure}
 The training dataset consists of two simulations of the flow past a cylinder for each Reynolds number in the range $Re = [90, 110, 130]$, resulting in six cases. For each case, $450$ probes are placed along the mean flow stream, uniformly distributed across the entire computational domain $\Omega$. Subsequently, $200$ of these probes are randomly removed, leaving a sparse set of $250$ probes. This sparse set of measurements of the mean flow $\mean\vdir$ is used as input to the GNN, while the output prediction is compared with the corresponding forcing stress tensor from the DNS ground truth. Fig.~\ref{fig:sparse_ground_curves}$a$ shows the positioning of the random probes on the mean flow, while Fig.~\ref{fig:sparse_ground_curves}$b$ shows the average training curves across the training dataset. 

In this case, we demonstrate an improvement in mean flow reconstruction across all the training cases, with an average improvement of $\mathcal{I} = 55.09\%$. This result highlights the robustness of the proposed approach in scenarios with sparse and randomly distributed measurements.

\subsection{Denoising}\label{subsec:denoising}

\begin{figure}[h]
    \centering
    \includegraphics[width=0.99\textwidth]{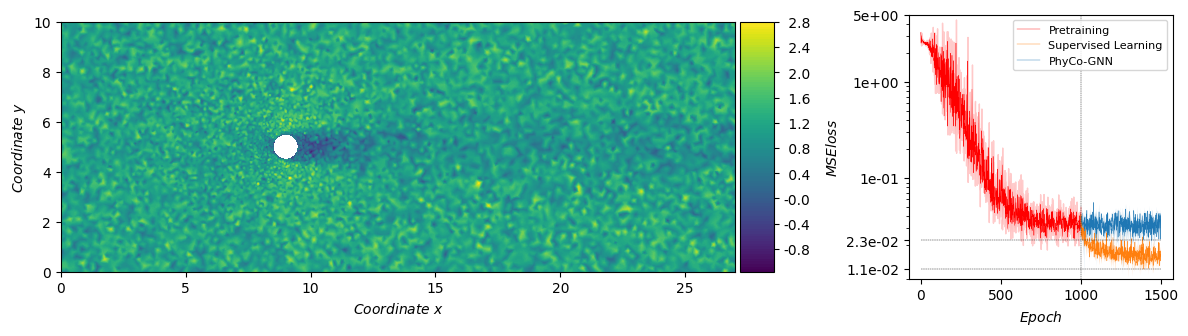}
    \includegraphics[width=0.99\textwidth]{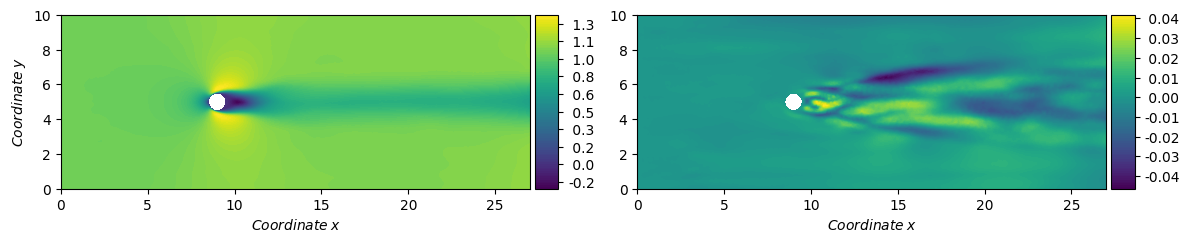}
    
    \begin{picture}(0,0)
        \put(-200, 216){$a)$}
        \put(  90, 216){$b)$} 
        \put(-200, 87){$c)$}
        \put(  25, 87){$d)$} 
    \end{picture}
    \vspace{-.5cm}
    \caption{Denoising -- training dataset: $3$ mean flow-forcing pairs at $Re = [90, 110, 130]$, perturbed with Gaussian noise; ($a$) mean flow at $Re = 130$; legend for ($b$), ($c$), and ($d$) as in Fig.~\ref{fig:baseline_150}}\label{fig:denoise_ground_curves}
\end{figure}

In this test case, the input mean flow field is perturbed with Gaussian noise. The probability density function used for the Gaussian distribution to generate the noise is given by:
\begin{equation}
    \psi\left(z\right) = \frac{1}{\sigma \sqrt{2\pi}}e^{\frac{-\left(z-\mu\right)^2}{2\sigma^2}}
    \label{eq:pdf_gaussian}
\end{equation}
where $z$ is the random variable, $\mu$ is the mean value of the normal distribution, and $\sigma$ represents its standard deviation. In this case, we assume $\mu = 0$, \emph{i.e.}, a standard normal distribution. The mapping operator $\mathcal{M}$, in this case, remains the identity operator.
The training dataset consists of three cases of cylinder flows at Reynolds numbers $Re = [90, 110, 130]$, each perturbed with Gaussian noise of fixed standard deviation $\sigma = 0.4$.
Fig.~\ref{fig:denoise_ground_curves}$a$ shows the effect of $\sigma = 0.4$ Gaussian noise on the mean flow (at $Re = 130$), while Fig.~\ref{fig:denoise_ground_curves}$b$ presents the accuracy of the mean flow reconstruction. The goal is to remove the Gaussian noise and accurately reconstruct the denoised mean flow field. Our approach demonstrates an improvement of $\mathcal{I} = 47.52\%$ on the training dataset, on average, across all cases.\\
Although not shown here, we also conducted additional tests at fixed Reynolds number ($Re = 130$), with varying noise levels $\sigma = [0.2, 0.4, 0.6]$. The results confirmed superior performance of the proposed approach compared to purely supervised learning, highlighting its robustness to different noise intensities.

\begin{figure}[t]
    \centering
    \includegraphics[width=0.99\textwidth]{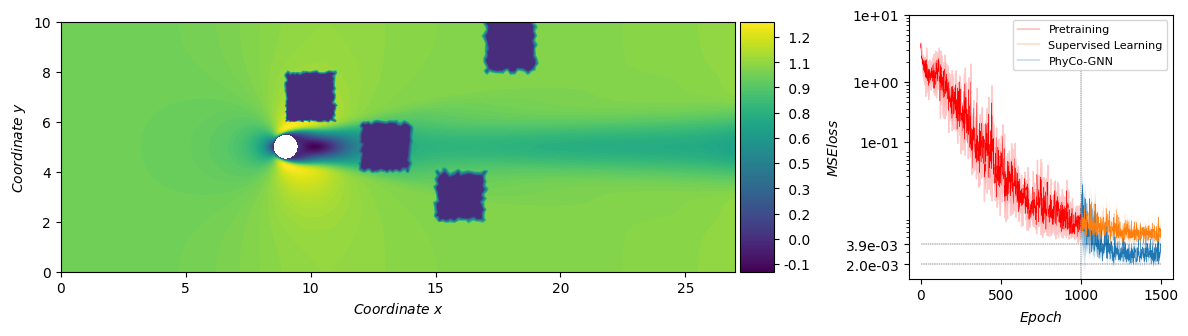}
    \includegraphics[width=0.99\textwidth]{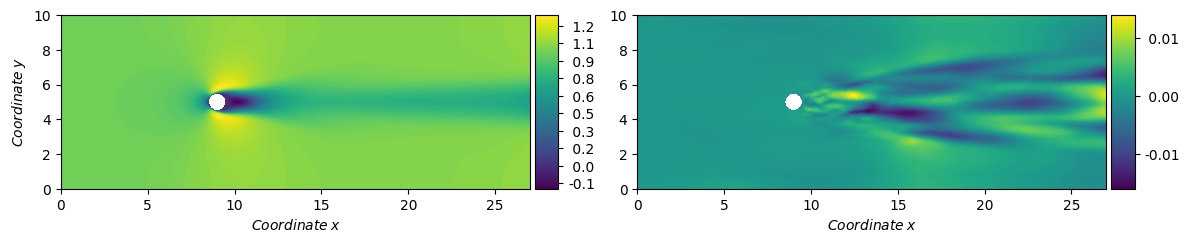}

    \begin{picture}(0,0)
        \put(-200, 216){$a)$}
        \put(  90, 216){$b)$} 
        \put(-200, 87){$c)$}
        \put(  25, 87){$d)$} 
    \end{picture}
    \vspace{-.5cm}
    \caption{Inpainting -- training dataset: $3$ mean flow-forcing pairs at $Re = [90, 110, 130]$, with randomly located patching masks; ($a$) mean flow at $Re = 110$; legend for ($b$), ($c$), and ($d$) as in Fig.~\ref{fig:baseline_150}}\label{fig:inpaiting_ground_curves}
\end{figure}
\subsection{Inpainting}\label{subsec:inpaiting}
In this test, masking patches are randomly applied to the input mean flow field. The training dataset consists of three cases of a cylinder obstacle at Reynolds numbers $Re = [90, 110, 130]$, each with different patch locations (Fig.~\ref{fig:inpaiting_ground_curves}$a$). In this setup, the mapping operator $\mathcal{M}$ used in the cost function excludes the masked regions, meaning that the comparison between the predicted and reference fields is performed only over the unmasked areas of the domain. The goal is to reconstruct the mean flow field by filling in the missing patches. The approach shows improvements on the training cases with an average increase of $\mathcal{I}=41.73\%$, successfully restoring the missing portions of the field and enhancing the overall reconstruction accuracy.

\medskip
A summary of the results is shown in Tab.~\ref{tab:results_comparison}, where the relative errors between the reconstructed mean flow and the ground truth are reported for both the pure supervised technique based on the GNN models (see also \cite{quattromini2023operator}) and the data-assimilation scheme PhyCo-GNN. The latter outperforms vanilla supervised learning in all cases as already shown in the previous paragraphs.

%
\input{TexFiles/table}

%% file: TexFiles/table.tex
\begin{table}[t]
\begin{center}
\resizebox{0.7\textwidth}{!}{
\begin{tabular}{|c|cc|}
\hline
\multirow{1}{*}{\textbf{Cases}} &\multicolumn{2}{c|}{\textbf{Reconstruction error}}
\\ \hline
&\multicolumn{1}{c|}{\textit{supervised}}             
&\multicolumn{1}{c|}{\textit{Phy-Co GNN}}
\\ \hline
{Reference case, Re=150}     
&\multicolumn{1}{c|}{0.0190} 
&\multicolumn{1}{c|}{\textbf{0.0126}}
\\ \hline
{Reference case, Re=90 (flip-flop)}     
&\multicolumn{1}{c|}{0.0910} 
&\multicolumn{1}{c|}{\textbf{0.0843}} 
\\ \hline
{Generalisation, Re=120}     
&\multicolumn{1}{c|}{0.0087} 
&\multicolumn{1}{c|}{\textbf{0.0069}} 
\\ \hline
{Sparse Measurements, Re=110}     
&\multicolumn{1}{c|}{0.0134} 
&\multicolumn{1}{c|}{\textbf{0.0092}}             
\\ \hline
{Gaussian Noise, Re=130}     
&\multicolumn{1}{c|}{0.0200} 
&\multicolumn{1}{c|}{\textbf{0.0103}}           
\\ \hline
{Inpainting, Re=110}     
&\multicolumn{1}{c|}{0.0056} 
&\multicolumn{1}{c|}{\textbf{0.0026}}         
\\ \hline
\end{tabular}}
\end{center}
\caption{Summary of the reconstruction errors for each test considered in Sec.~\ref{sec:results}. The cases listed correspond to those presented in Figures \ref{fig:baseline_150}--\ref{fig:inpaiting_ground_curves}. For each case, a comparison is made between the baseline method based on supervised learning and the performance of PhyCo-GNN, with the errors reported as relative errors in the $2$-norm.} \label{tab:results_comparison}
\end{table}

%% file: TexFiles/9_conclusion.tex
\section{Conclusion}\label{sec:conclusion}

In this work, we introduced a direct-adjoint data assimilation scheme based on the Reynolds-Averaged Navier-Stokes (RANS) equations, where the closure model is based on Graph Neural Networks (GNNs). The resulting scheme, named Physics-Constrained Graph Neural Network (PhyCo-GNN), demonstrated the potential of combining machine learning techniques with well-established physical principles to improve the accuracy and reliability of mean flow reconstruction. More specifically, the forcing term of the RANS equations is modelled using a GNN model informed by gradients obtained through auto-differentiation of the neural network and gradients computed analytically based on the adjoint equations associated with the iterative process. On one hand, these gradients ensure that the learned model adheres to the underlying physics. On the other hand, the closure model benefits from the flexibility of GNNs in handling unstructured meshes, making the framework particularly well-suited for complex geometries often encountered in Computational Fluid Dynamics (CFD). Finally, the combination of GNN and adjoint-based methods mitigates the dependency on large datasets, as the inclusion of physics reduces the need for extensive training data to achieve reliable predictions.

We tested the PhyCo-GNN framework across several scenarios, with a particular focus on reconstructing the mean flow from sparse measurements, noisy probes, and incomplete flow fields (inpainting). The models are compared with a supervised learning method, which serves as a benchmark, where the learning process of the GNN model is not constrained by the adjoint but solely relies on numerical data. In particular, we consider as the reference case the mean flow reconstruction of flows past a 2D cylinder at a Reynolds number of $Re = 150$ and a two-cylinder configuration at $Re = 90$. Other tests include generalisation by considering interpolation and extrapolation with respect to unseen Reynolds numbers during the training process, reconstruction from noisy inputs, and inpainting, which involves the reconstruction of the mean flow field by filling in the missing patches. For all test cases, the PhyCo-GNN approach showed substantial improvements compared to the reference method based on pure supervised learning.

Future work will explore the extension of the presented framework to more complex scenarios, including higher Reynolds number regimes, three-dimensional flows, and experimental datasets, which are often characterised by sparse, noisy, or incomplete measurements. While the methodological structure of the PhyCo-GNN framework is, in principle, compatible with such configurations, the transition to large-scale 3D domains poses significant computational challenges. These include the increasing cost of mesh resolution, memory usage during GNN training, and the resource demands of adjoint-based PDE solvers. Addressing these limitations will require adopting scalable strategies such as domain decomposition for graph-based learning (a review can be found in Dolean \emph{et al.} \cite{dolean2024}), modular and multi-scale GNN architectures, and multi-GPU distributed training pipelines. In parallel, the adoption of more efficient FEM backends and the exploration of alternative network architectures, such as physics-informed transformers or recurrent graph networks, may further enhance the expressiveness and applicability of the framework.

%% file: TexFiles/10_appA.tex
\section{Hyperparameters}\label{appendixB:hyperparameter}

The performance of a neural network is determined by its hyperparameters, which are set before training and influence both the model's architecture and the training process. These parameters define the network's ability to approximate complex functions, balance computational efficiency, and ensure stable training.

To identify the optimal set of hyperparameters, we used the open-source optimisation library \verb|Optuna| \citep{akiba2019optuna}. Optuna combines efficient parameter space exploration with advanced pruning strategies, dynamically pruning unpromising trials to reduce computational costs while maximising performance. The final hyperparameters selected for the learning tasks in this study are:
\begin{itemize}
    \item \underline{Embedded dimension}: $d_h = 35$. The size of the hidden feature space for each node in the GNN. A higher dimension enables richer representations of node features, but excessively large values increase computational cost without necessarily improving performance.
    
    \item \underline{Number of GNN layers}: $k = 40$. This depth ensures global information can propagate across the graph, while avoiding over-smoothing.
    
    \item \underline{Update relaxation weight}: $\alpha = 6 \times 10^{-1}$. This parameter balances the influence of new and old information during feature updates.
    
    \item \underline{Loss function weight}: $\gamma = 0.1$. Controls the weighting of loss terms from different layers in the GNN during training. Higher weights on later layers emphasise long-range dependencies in the training process.
    
    \item \underline{Learning rate}: $LR = 3 \times 10^{-3}$. This optimises the step size for weight updates, balancing convergence speed and stability.
\end{itemize}
These hyperparameters were optimised to achieve a balance between model accuracy and computational efficiency. It is important to note that these hyperparameters are task-dependent and, in theory, should be re-optimised for each specific learning task to ensure optimal performance. However, re-optimisation for the different learning tasks in this study showed that the hyperparameters remained largely consistent. This suggests that the GNN architecture exhibits inherent robustness to variations in hyperparameters within the range of learning tasks addressed here.